\documentclass[aps,prl,superscriptaddress,twocolumn,tightenlines]{revtex4-1}
\usepackage[pdftex]{graphicx,xcolor}
\usepackage{amsmath}
\usepackage{amsfonts}
\usepackage{amssymb}
\usepackage{mathrsfs}
\usepackage{bbm}
\usepackage{bbold}
\usepackage{subfigure}

\newcommand {\be}{\begin{eqnarray}}
\newcommand {\ee}{\end{eqnarray}}

\begin{document}
%\begin{fmffile}{fmdiag}

\title {Dimer Impurity Scattering, Reconstructed Nesting and Density-Wave Diagnostics in Iron Pnictides}
\author {Jian Kang}
%\affiliation {Institute for Quantum Matter and Department of Physics \& Astronomy, The Johns Hopkins University, Baltimore, MD 21218}
\author {Zlatko Te\v sanovi{\' c}}
\affiliation {Institute for Quantum Matter and Department of Physics \& Astronomy, The Johns Hopkins University, Baltimore, MD 21218}
\date {\today}

\begin{abstract}

While the impurity-induced nanoscale electronic disorder has been extensively reported in the underdoped iron pnictides, its microscopic origins remain elusive. Recent scanning tunneling microscopy (STM) measurements reveal a dimer-type resonant structure induced by cobalt doping. These dimers are randomly distributed but uniformly {\em aligned} with the antiferromagnetic $a$ axis. A theory of the impurity-induced quasiparticle interference patterns is presented that shows the local density of states developing an oscillatory pattern characterized by both geometry and
orbital content of the {\em reconstructed} Fermi pockets, occasioned by the pocket density-wave (PoDW) order
along the $b$ axis. This pattern breaks the $C_4$ symmetry and its size and orientation
compare well with the dimer resonances found in the STM experiments, hinting at the presence of a ``hidden" PoDW order. More broadly, our theory spotlights such nanoscale structures as a useful diagnostic tool for various forms of order in iron pnictides.

\end{abstract}
\maketitle

%\pacs{74.25.Jb, 74.25.Qt, 74.72.-h }

The iron pnictide high-$T_c$ superconductors \cite{LaOFeAs,BaFe2As2} exhibit several remarkable features \cite{Greene}. Among them are the proximity and interplay of antiferromagnetism (AFM) and a structural transition \cite{ST1111,ST122}, manifested by anisotropy in electronic properties \cite{ResAnisotropy,ResNematic,EleAnisFeAs,NematicSTM2010,OrbAnis122,EleAnis111}.
Various experiments reveal strong correlation between
these two transitions \cite{STMTLinear, STMTTrOrder}.
A prevalent explanation is that the structural transition results from the fluctuations of incipient AFM order \cite{STMTCoupling, NemMagFluc}. Alternatives include the key role of orbital degrees of freedom \cite{Phillips} and, in particular, the proposal that the structural transition originates from the pocket density-wave (PoDW) \cite{VDWJian} in parent compounds, a ``hidden" order responsible for orbital ferromagnetism.

Recent scanning tunneling microscopy (STM) experiments observe the nematic-type electronic formations developing around dopant atoms \cite{NematicSTM2010,DavisDimer}. These formations appear as ``dimer resonances," with two neighboring peaks separated by $\sim 6-8$ lattice spacings.
Importantly, the dimers are oriented {\em along} the $a$ axis of the pnictides'
unidirectional AFM order.
In this Rapid Communication, we show that both the appearance of dimer resonances and their size and orientation provide a direct insight into the electronic structure and correlations in iron pnictides. In this regard, the properties of such dimers can serve as a diagnostic tool to unravel the nature of the underlying
microscopic ground state and the sequence of Fermi-surface reconstructions left in the wake
of various itinerant density-wave (DW) orderings \cite{Hoffman}.

These are our main results: The dimer resonance \cite{NematicSTM2010,DavisDimer} is a consequence of a type of ``reconstructed" nesting \cite{ReconNestARPES} characterized by the wavevector $\vec q_a \approx (0.4\pi, 0)$. Such
nesting tendency is manifest in our detailed calculations
within the three orbital model \cite{DaghoferMod},
where $\vec q_a$ -- associated with the short axis of the elliptical electron pocket --
emerges as a prominent feature of the reconstructed band structure,
itself occasioned by the transition from the paramagnetic phase to the PoDW. Specifically:
{\em i)}~In the paramagnetic phase, the Fermi pocket is far from
any nesting at $\vec q_a$. Consequently, no resonances appear;
{\em ii)}~As a PoDW is formed, the electron pocket is itself deformed while,
simultaneously, sections of the reconstructed hole pockets become flatter
as the PoDW order parameter increases. This gives rise to
the ``reconstructed" nesting at $\vec q_a$;
{\em iii)}~$\vec q_a$ is {\em perpendicular}
to the PoDW, producing the real-space dimer pattern breaking the $C_4$ symmetry.
The resonant pattern arises when randomly distributed
dopant impurities induce two peaks in the local density of states (LDOS), {\em both their separation
and direction set by} $\vec q_a$;
{\em iv)}~When a PoDW and a SDW coexist, two
nesting vectors $\vec q_a \approx (0.4\pi, 0)$ and $\vec q_b \approx (0, 0.4\pi)$ emerge from the reconstructed pockets. With $e_y$ and $e_x$ coupled to the inner and outer hole pockets, $h_2$ and $h_1$, respectively, the susceptibility near $\vec q_a$ remains much larger than its counterpart
near $\vec q_b$. Again,
the $C_4$ symmetry is broken resulting in dimer patterns observed in
\cite{NematicSTM2010,DavisDimer}.

We now supply the details behind the above physical picture.
Experiments and theoretical calculations indicate that iron
pnictides contain four disconnected Fermi-surface pockets (Fig. 1) \cite{VladBand} \cite{RichardARPES}.
Additional parts of the Fermi surface are present
 in some materials \cite{RichardARPES} but this is not
important for the physics discussed here.
When two hole pockets closely match two electron pockets,
as is the case in many iron pnictide parent compounds \cite{RichardARPES},
the geometric nesting favors the formation of
two DWs \cite{VDWJian}: The PoDW partially gaps $e_y$ and one hole pocket and induces a structural transition,
while $e_x$ and the remaining hole pocket form the partially gapped SDW.
The theory \cite{VDWJian} naturally explains the proximity of the two transitions and
accounts for the observed orbital ferromagnetism.

While geometric nesting tendencies are important \cite{VladBand}, they are not the complete story:
The orbital content of the Fermi pockets must be considered as well \cite{CobaltVHirs}.
Here, we start with a simplified geometric model allowing for an analytic glance at the
physics and then fortify the results within a detailed numerical follow-up
employing three-orbital content \cite{DaghoferMod}, known to capture the
main features of real materials \cite{RichardARPES}.
We model the Fermi pockets as:
\be
\epsilon &_{h_1, h_2}({\vec k}) =\epsilon_0 - \frac{k^2}{2m} \nonumber
\ee
\be
\epsilon_{e_x}({\vec k}) = \frac{(k_x - \pi)^2}{2m_a} + \frac{k_y^2}{2m_b} - \epsilon_0 ~;~\epsilon_{e_y} = \epsilon_{e_x}(k_x\leftrightarrow k_y)
\label{SimMod}
%& = & \frac{k_x^2}{2m_b} + \frac{(k_y - \pi)^2}{2m_a} - \epsilon_0
\ee
with $m_a  > m  > m_b$, so that the two DWs are partially gapped when order parameters $\Delta_{PoDW}$ and $\Delta_{SDW}$ are small. $e_y$ and $e_x$
bands are related by $90^{\circ}$ rotation. To simplify the analytic calculation, we assume that $h_1$ and $h_2$ are isotropic and have the same dispersion relations.

\begin{figure}[htb]
\includegraphics[scale=0.35]{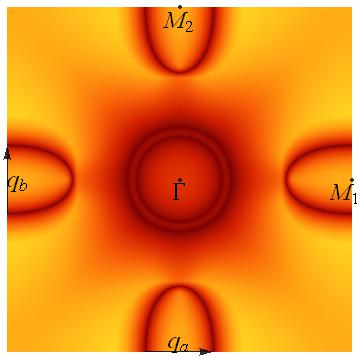}
\caption{(Color online) Fermi pockets in the unfolded Brillouin zone of iron pnictides.
Two hole pockets $h_1$ and $h_2$ are centered at the $\Gamma = (0 , 0)$ point.
The elliptical electron pockets $e_x$ and $e_y$ are centered at
 $\vec M_1 = (\pi, 0)$
and $\vec M_2 = (0, \pi)$, respectively. $h_1$, $h_2$, $e_x$, and $e_y$ pockets exhibit
strong nesting tendencies in many parent compounds. Here, we propose
that the resonant electronic structure arises due to the ``reconstructed" nesting with the wave vector $\sim\vec q_a$, the shorter axis of the $e_y$ pocket.}
\label{Fermi_Pocket_UnBZ}
\end{figure}

It is known that static charge/spin susceptibilities contain peaks around the
nesting vectors $\vec M_1=(\pi, 0)$ and $\vec M_2 = (0, \pi)$ in
the above noninteracting model, implying a tendency
to DW formation in moderately correlated iron pnictides \cite{VladBand}. Initially, by ignoring the orbital
content at the Fermi level within \cite{DaghoferBand}, we can determine the
analytic form for such geometric susceptibility at $T=0$, within
a {\em single} elliptical electron pocket (\ref{SimMod}):
%\[ \epsilon(\vec k) = \frac{k_x^2}{2 m_a} + \frac{k_y^2}{2 m_b} - \epsilon_0 \]
%for the ground state ($T  = 0$). The formula is
\begin{eqnarray*}
  & & \chi(q_x, q_y) = 2 \int \frac{\mathrm{d}^2 k}{(2\pi)^2} \frac{n_f(\epsilon(\vec k)) - n_f(\epsilon(\vec k - \vec q))}{\epsilon(\vec k - \vec q) -\epsilon(\vec k) + i 0^+} \\
  & = & \left\{ \begin{array}{cc}
    \frac{\sqrt{m_a m_b}}{4 \pi} & \mbox{if } \frac{q_x^2}{2 m_a} + \frac{q_y^2}{2 m_b} < 4 \epsilon_0 . \\
    \frac{\sqrt{m_a m_b}}{4 \pi} \left[ 1 - \sqrt{1 - \frac{4 \epsilon_0}{\frac{q_x^2}{2 m_a} + \frac{q_y^2}{2 m_b}} }  \right]  & \mbox{if } \frac{q_x^2}{2 m_a} + \frac{q_y^2}{2 m_b} > 4 \epsilon_0 .
  \end{array}  \right.
\end{eqnarray*}
$\chi({\vec q})$ is effectively a constant for $\vec q/2$ within the pocket, and decreases slowly as $\vec q/2$ moves outside. Consequently, $\chi(\vec q)$ contains only an unremarkable
ridge instead of a peak,
implying no resonance around $\vec q_a$ in the normal state of such a simple geometric model.
These features of geometric nesting are echoed in a
realistic calculation with full orbital content \cite{VladBand, DaghoferBand}:
%\begin{widetext}
  \begin{eqnarray}
  & &\chi(\vec r, \vec r') =  \langle \rho(\vec r) \rho(\vec r') \rangle   =  \sum_{\alpha, \beta} \langle d_{\alpha}^{\dag} (\vec r) d_{\alpha} (\vec r)  d_{\beta}^{\dag} (\vec r') d_{\beta} (\vec r')   \rangle \nonumber  \\
  & = & -\frac2{N^2} \sum_{k_i, \mu_i}  \langle d_{\mu_1}^{\dag}(\vec k_1)  d_{\mu_4}(\vec k_4)  \rangle \langle d_{\mu_3}^{\dag} (\vec k_3) d_{\mu_2} (\vec k_2) \rangle \langle \mu_1, \vec k_1 \vert \mu_2, \vec k_2 \rangle \nonumber \\
   & & \langle  \mu_3, \vec k_3  \vert \mu_4, \vec k_4 \rangle \exp\left( i (\vec k_2 - \vec k_1)\vec r + i(\vec k_4 - \vec k_3)\vec r'  \right)~,
\label{Sus_Def}
\end{eqnarray}
%\end{widetext}
where $\alpha , \beta$ denote $d$ orbitals and $\mu_i$s are the band indices.
In the paramagnetic phase, the Green's function
$\langle d_{\mu_i}(\vec k_i,\omega_n)  d^{\dag}_{\mu_j}(\vec k_j,\omega_n)\rangle$ is finite only if $\vec k_i = \vec k_j$ and $\mu_i = \mu_j$ and thus Eq.~(\ref{Sus_Def}) gives
\be
 \chi(\vec q) = \frac2N \sum_{\vec k, \mu, \nu} \frac{\left\vert \langle \mu, \vec k + \vec q  \vert \nu, \vec k \rangle \right\vert^2 \left( n_f(\epsilon_{\nu}) - n_f(\epsilon_{\mu}) \right) }{\epsilon_{\mu}(\vec k + \vec q) - \epsilon_{\nu}(\vec k) + i 0^+}.
 \label{Sus_Disorder}
\ee
The factor of 2 in (\ref{Sus_Def}) is due to spin and $n_f(\epsilon)$ is the Fermi distribution function. Again, we find nesting peaks at $(\pi, 0)$ and $(0, \pi)$ but no enhanced structure near $\vec q_a$.

\begin{figure}[htb]
\includegraphics[scale=0.50]{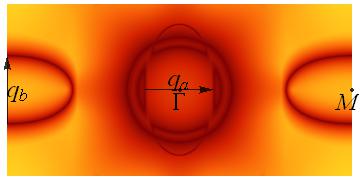}
\caption{(Color online) As PoDW order develops, the Fermi surface reconstructs, and
$e_y$ and the inner hole pocket $h_2$ are deformed. The curvature along
$e_y$ decreases, leading to the ``reconstructed" nesting at $q_a$, the short axis of the $e_y$ pocket. This reconstructed nesting is absent in the paramagnetic phase, but becomes
pronounced as $\Delta_{PoDW}$ increases.}
\label{Fermi_Pocket_PDW}
\end{figure}

The situation changes once the DW order is established. Within a PoDW, which
is the leading DW instability of Ref. \cite{VDWJian}, the electron pocket $e_y$ couples with one
of the hole pockets; in the following, we assume this is $h_2$. The
reconstructed Fermi surface is shown in Fig.~\ref{Fermi_Pocket_PDW}.
%\begin{figure}[htb]
%\includegraphics[scale=0.35]{Fermi_Pockets.eps}
%\includegraphics[scale=0.50]{ChiSimple.jpg}
%\caption{Charge susceptibility at $\vec q = (q_a, 0)$ as a function of $\Delta_{PDW}$ for the simple model in eqn~\ref{Sim_Mod}.
%It can be seen that the susceptibility increases as $\Delta_{PDW}$ becomes larger, because part of the $e_y$ pocket becomes %straighter. This creates ``reconstructed" nesting with the wave vector of $q_a$.}
%\label{Fermi_Pocket_UnBZ}
%\end{figure}
$\chi(\vec q_a)$ now increases, reflecting the nesting between the {\em reconstructed} electron pockets: $e_y$ deformation promotes $\vec q_a$ to a ``reconstructed" nesting vector, connecting separate portions of two small $e$ pockets, as depicted in Fig. \ref{Fermi_Pocket_PDW}. When $\Delta_{PoDW}$ is small, $\vec q =\vec q_a$ is a local maximum of $\chi(\vec q)$, with a strong peak when the nesting is optimized, i.e. $\bigl (\partial^2 \lambda /\partial k_y^2\bigr )_{k_x = \pm q_a/2, k_y = 0} = 0$, $\lambda$ being the energy of the reconstructed $h$ pocket.
Within our simple model (\ref{SimMod})
\be
  \Delta_{opt} = \epsilon_0 \frac{m - m_b}{m_a + m_b} \sqrt{\frac{m_a}{m}}
  \label{Opt_Nest}
\ee
optimizes nesting. For $\Delta_{PoDW} \ll \Delta_{opt}$, $\chi$ increases by
\be
 & & \delta \chi(\vec q_a)  =  \chi_{PoDW}(\vec q_a) - \chi_{para}(\vec q_a) \nonumber \\
 & \approx & \frac{\sqrt{2 m_a m_b}}{\pi^2} \left( \frac{m_a - m}{m_a - m_b}  \right)^{\frac14} \left( \frac{\Delta}{\epsilon_0} \right)^2 \frac{m (m + m_a)}{(m - m_b)^2}~,
\ee
while, for $\Delta_{PoDW} = \Delta_{opt}$,
\begin{eqnarray}
 \delta \chi(\vec q_a) & \approx & \frac1{\pi^2} \left( \frac{\Delta k}{v_F^3 \beta} \right)^{1/4} \label{Delta_Sus_PDW} \\
 \Delta k & \approx & \sqrt{\frac{\epsilon_0}{m + m_b}} \frac{m_a - m_b}{m_a + m_b} (m - m_b) \label{deltak} \\
 v_F & = & \sqrt{\frac{2 \epsilon_0}{m_b}} \frac{m_a - m_b}{\sqrt{(m + m_a)(m_a + m_b)}} \label{vf_recon} \\
 \beta & =  & \left( \frac1m + \frac1{m_a} \right) \frac{m_a + m_b}{32 \epsilon_0 m_b (m - m_a)} \nonumber \\
 & & \left[ 1- \frac14 \left( \frac{m_a - m}{m + m_b} \right)^2  \right] \label{beta}~.
\end{eqnarray}
$\Delta k$ is width of the reconstructed $h$ pocket, $v_F$ is the Fermi velocity $\parallel \hat x$, $\vec k = \vec q_a/2$, and
$\beta = \bigl(\partial^4 \epsilon/\partial k_y^4\bigr)_{k_x = q_a/2, k_y = 0}$.

\begin{figure}[htb]
\centering
\vspace{-0.25cm}\includegraphics[scale=0.32]{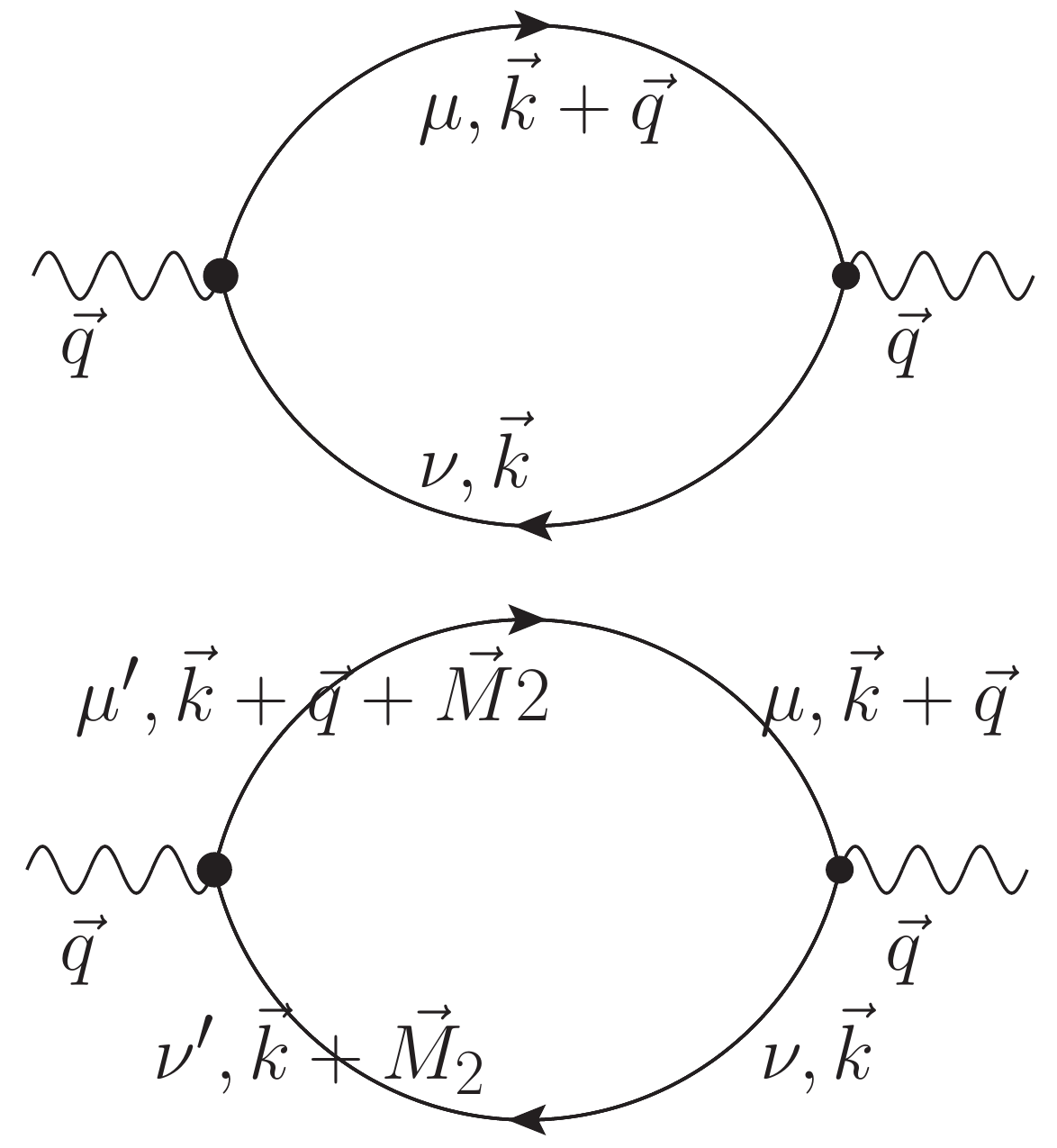} \hspace{0.03cm}
\includegraphics[scale=0.35]{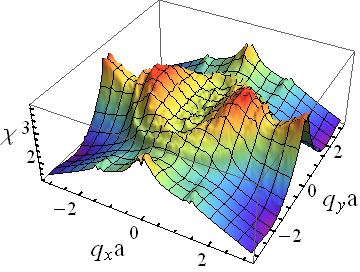}
\caption{(Color online) Left: Feynman diagrams for $\chi^c$ in the PoDW phase. All other diagrams cancel out.
Right: $\chi^c(\vec k)$ with $\Delta_{PoDW} = 20 meV$ in the realistic three-orbital model. The original peaks at $(0, \pm \pi)$ in the paramagnetic phase are suppressed by finite $\Delta_{PoDW}$. Instead, $\chi(\vec k)$ contains a plateau centered at $\vec q_a \approx (\pm 0.4 \pi, 0)$,
breaking the $C_4$ symmetry. This plateau arises from
the ``reconstructed" nesting at $q_a$.}
\label{Sus_PDW_Plot}
\end{figure}

Again, we fortify the above picture with the realistic calculation of charge susceptibility $\chi^c(\vec q)$
within the three orbital model. In the PoDW state, $\langle d_{\mu_i}(\vec k_i,\omega_n) d^{\dag}_{\mu_j}(\vec k_j,\omega_n) \rangle$ is finite only if
$     \mu_i = \mu_j$ and $\vec k_i = \vec k_j $ or
 $ (\mu_i, \mu_j) = (e_y, h_2)$ and $\vec k_i = \vec k_j + (0, \pi)$.
%\be
 %  \left\{ \begin{array}{l}
%     \mu_i = \mu_j \mbox{ and } \vec k_i = \vec k_j \qquad \mbox{or} \\
%     (\mu_i, \mu_j) = (e_y, h_2) \mbox{ and } \vec k_i = \vec k_j + (0, \pi)
 %  \end{array}       \right.  .
  % \label{GreenFunc_PDW}
%\ee
Figure \ref{Sus_PDW_Plot} displays Feynman diagrams contributing to $\chi^c(\vec k)$. Additional
diagrams, composed of one normal and one anomalous Green's function, break time reversal symmetry and cancel out upon summation over $\vec k$. $\chi^c$ is calculated within the three-orbital model
and is shown in Fig.~\ref{Sus_PDW_Plot}. Evidently, a high plateau around $(\pm 0.4\pi, 0)$ arises due to the ``reconstructed nesting" at vector $\vec q_a$. This is qualitatively consistent with the results in Eq.~(\ref{Delta_Sus_PDW}) for the simple model (\ref{SimMod}). In addition, the original peaks at $\vec M_2 = (0, \pm \pi)$ are suppressed due to the formation of a PoDW.

At low $T$, Ref. \cite{VDWJian} predicts the coexistence of a PoDW and a SDW, and therefore, leads to two ``reconstructed" nesting vectors, $\vec q_a$ and $\vec q_b$, related by $90^{\circ}$ rotations. If $\Delta_{PoDW} \neq \Delta_{SDW}$, the $C_4$ symmetry is already broken by this unequal pairing.
In Ref.~\cite{VDWJian}, $\Delta_{PoDW} > \Delta_{SDW}$ and this naturally leads
to $\chi(\vec q_a) > \chi(\vec q_b)$. When $\Delta_{PoDW} = \Delta_{SDW}$, however, the difference between $\chi(\vec q_a)$ and $\chi(\vec q_b)$ could still arise from the distinct orbital
content of Fermi pockets. To illustrate the effect of orbital content, we assume
\cite{RichardARPES}
\begin{eqnarray}
  h_1 & = & \cos(\theta) d_{yz} + \sin(\theta) d_{xz} \\
  h_2 & = & \cos(\theta) d_{xz} - \sin(\theta) d_{xz} \\
  e_x & = & d_{yz} \quad  e_y = d_{xz}~,
\end{eqnarray}
where $\theta$ is the polar angle at the Fermi pocket.

\begin{figure}[htb]
\includegraphics[scale=0.35]{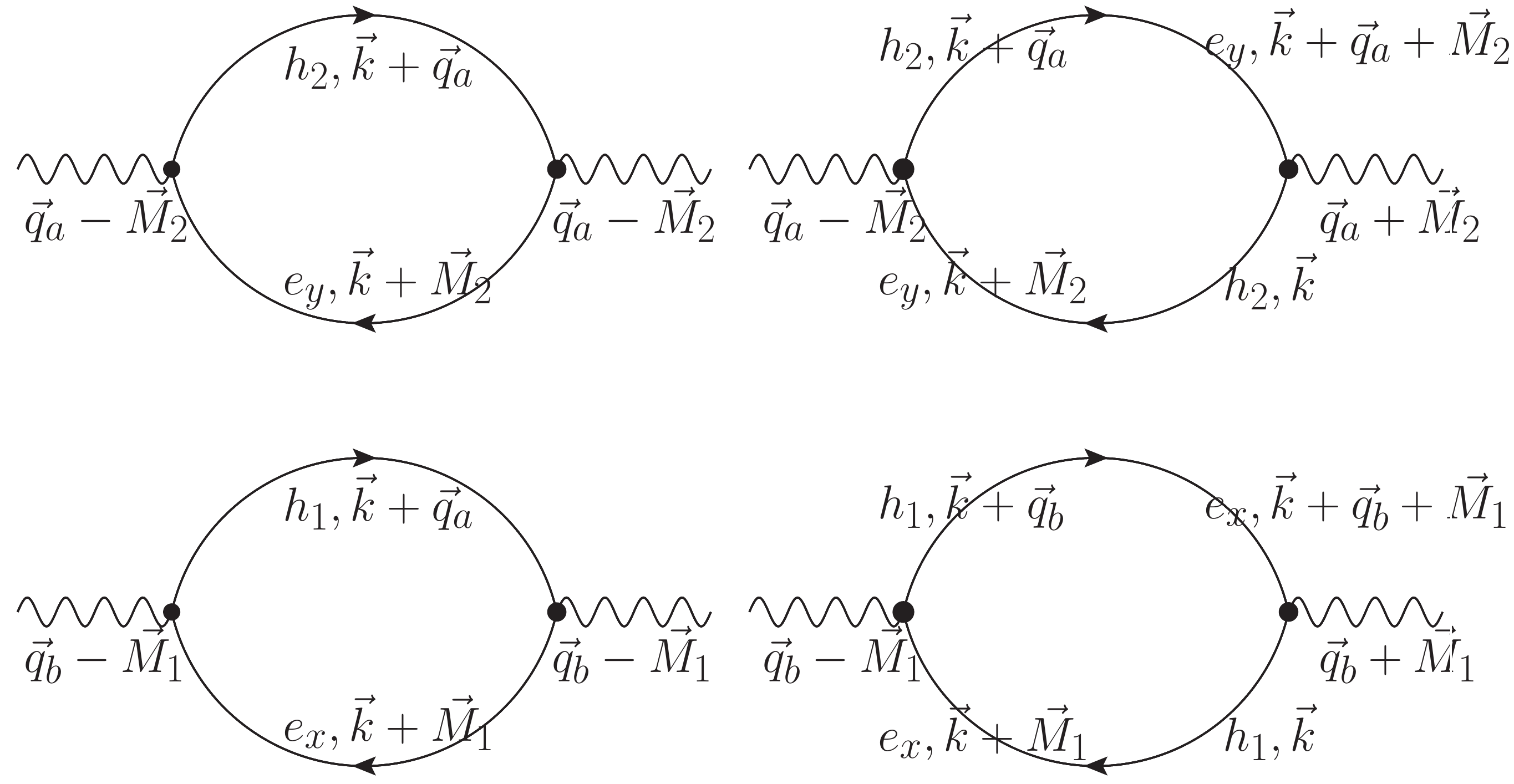}
\caption{$C_4$ symmetry is broken due to orbital content. Only the most significant diagrams contributing to the difference between $\chi(\vec q_b)$ and $\chi(\vec q_a)$ are displayed. The
upper and lower two contribute to $\chi(\vec q_a)$ and $\chi(\vec q_b)$, respectively. This difference arises from the vertex, which includes orbital overlaps between incoming and outgoing
fermions. The diagrams which have $C_4$ symmetric partners are not shown.}
\label{C4Broken_Feyn}
\end{figure}

Figure.~\ref{C4Broken_Feyn} shows the Feynman diagrams contributing to the susceptibility difference $\delta \chi = \chi(\vec q_a + \vec M_2) - \chi(\vec q_b + \vec M_1)$. Quantitatively, $\delta \chi$ is dominated by the first diagram:
\begin{eqnarray}
  \delta \chi & \approx  & \int \frac{\mathrm d \vec k_1}{(2\pi)^2} (\cos^2 \theta(\vec k + \vec q) - \sin^2 \theta(\vec k + \vec q)) \nonumber \\
  & & u^2(\vec k + \vec q) v^2(\vec k) \frac{n_f(\lambda (\vec k)) - n_f(\lambda (\vec k + \vec q)) }{\lambda(\vec k + \vec q) - \lambda (\vec k)}.
\end{eqnarray}
The vertex factors $\cos^2 \theta(\vec k + \vec q)$ and  $\sin^2 \theta(\vec k + \vec q)$
come from the orbital component of the charge density $\rho(\vec r)$ when calculating $\chi(\vec q_a)$
and $\chi(\vec q_b)$, respectively,
%$u(\vec k)$ and $v(\vec k)$ are coherence factors:
\begin{eqnarray*}
  u^2(\vec k) & = & \frac12\left( 1 + \frac{\left( \epsilon_h(\vec k) - \epsilon_e(\vec k + \vec M) \right)/2}{\sqrt{\left( ( \epsilon_h(\vec k) - \epsilon_e(\vec k + \vec M) )/2  \right)^2 + \Delta^2}}  \right) , \\
  v^2(\vec k) & = & 1 - u^2(\vec k)~,
\end{eqnarray*}
$n_f$ is the Fermi function, and $\lambda$ is the energy of the reconstructed, partially gapped fermions. The other diagrams give similar contributions, with different vertices.

It is now difficult to get an analytic expression for $\delta \chi$ even within the model (\ref{SimMod}).
However, for $\Delta_{PoDW} = \Delta_{SDW} = \Delta_{opt}$ in (\ref{Opt_Nest}) and at $T=0$,
%$\delta \chi$ is well approximated by
\be
\delta \chi \approx \frac1{\pi^2} \left( \frac{\Delta k}{v_F^3 \beta} \right)^{\frac{1}{4}}  \frac{\sqrt{m_a m_b (m - m_b) (m_a + m)}}{m(m_a + m_b)},
\ee
with $\delta k$, $v_f$ and $\beta$ from Eqs. (\ref{deltak})--(\ref{beta}).
The anisotropy is determined by the choice of coupling between the $e$ and $h$ pockets:
$\chi(\vec q_a + \vec M_2) > (<) \chi(\vec q_b + \vec M_1)$ when $e_y$ and $e_x$ couple to
$h_2$($h_1$) and $h_1$($h_2$), respectively
\cite{footnotereverse}.
%, respectively. Conversely, if $e_y$ couples to , and $e_x$ to , then $\chi(\vec q_a + \vec M_2) < \chi(\vec q_b + \vec M_1)$.

\begin{figure}[htb]
\includegraphics[scale=0.50]{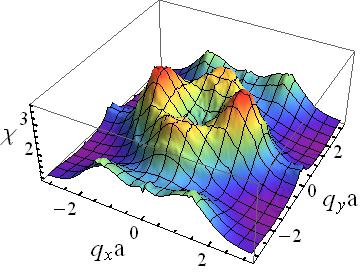}
\caption{(Color online) $\chi^c(\vec k)$ with $\Delta_{PoDW} = 20 meV$ and $\Delta_{SDW} = 15 meV$ within
the three-orbital model. The original peaks around $(0, \pm \pi)$ and $(\pm \pi, 0)$ in the paramagnetic phase are suppressed due to finite $\Delta_{PoDW}$. In addition, two small peaks emerge around $(\pm 0.4\pi, \pm\pi)$, breaking the $C_4$ symmetry. }
\label{Sus_Coexist_Plot}
\end{figure}
Again, we follow up with the realistic calculation of $\chi^c(\vec q)$ based on the three orbital model.
The results are displayed in Fig. \ref{Sus_Coexist_Plot}. The peaks at the nesting vectors $\vec M_1$ and $\vec M_2$ are suppressed due to the PoDW and SDW. Evidently, $\chi(\vec k)$ breaks the $C_4$ symmetry, as expected.

STM experiments reveal an anisotropic electronic dimer structure developing when the iron pnictide is doped with cobalt \cite{DavisDimer}. Our theory of reconstructed nesting can account for this
phenomenon, by considering the impurity-induced quasiparticle interference in the parent compounds.
First, the local potential of the cobalt atom is
$H_{imp} = \sum_{\sigma, \alpha} (V^{\alpha}_s + \sigma V^{\alpha}_m) d_{\alpha}^{\dag}(\vec r) d_{\alpha}(\vec r)$,
%\be
%H_{imp} = \sum_{\sigma, \alpha} (V^{\alpha}_s + \sigma V^{\alpha}_m) d_{\alpha}^{\dag}(\vec r) d_{\alpha}(\vec r)~,
%\label{H_Imp}
%\ee
where $V_s$ and $V_m$ are the nonmagnetic and the magnetic parts of the impurity potential,
respectively, and are given for each orbital $\alpha$ in
\cite{CobaltVHirs}. The LDOS with a single impurity is:
%which we determine by T-matrix method:
\begin{eqnarray}
  \rho(E, \vec r') & = & -\frac1{\pi} Im \sum_{\alpha, \sigma} G_{\alpha \alpha, \sigma}(i \omega \longrightarrow E + i 0^+, \vec r', \vec r') \nonumber \\
  G_{\sigma} (i \omega, \vec r', \vec r') & = & G_0(\vec r', \vec r') + G_0(\vec r', \vec r) V_{imp} G_0(\vec r, \vec r') \\
  V_{imp} & = & H_{imp} (1 - G_0 H_{imp})^{-1} \label{Vimp}~;
\end{eqnarray}
$G_0$ is the bare Green's function matrix in orbital space.
%dependent only on $\vec r_1 - \vec r_2$.
%and $H_{imp}$ in (\ref{Vimp}) is given by (\ref{H_Imp}).

\begin{figure}[htb]
\includegraphics[scale=0.50]{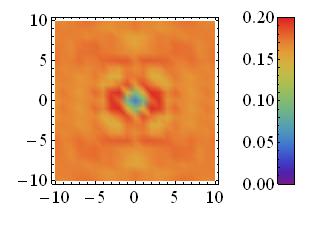}
\caption{(Color online) The LDOS integrated from $E=0$ to $37meV$ \cite{DavisDimer},
with $\Delta_{PoDW} = 20meV$, $\Delta_{SDW} = 15meV$,
and the impurity located at $\vec r =(0,0)$. We assume $e_y$ ($e_x$) couples to $h_2$ ($h_1$). LDOS peaks at $\sim\vec r = (\pm 3a, 0)$, fixing the dimer size to $6a$. The dimers are aligned with
the $x$ axis, reflecting the orientation of the peaks of charge susceptibility
(Fig. \ref{Sus_Coexist_Plot}). This anisotropy is due to different magnitudes of DW order parameters and different orbital content of individual pockets.}
\label{LDOS}
\end{figure}

Figure.~\ref{LDOS} shows the LDOS on a $10 \times 10$ square lattice, centered around the cobalt impurity. Note the pronounced peaks at $\vec r = (\pm 3 a, 0)$,
giving rise to the anisotropic dimer structure, observed in \cite{DavisDimer}. The
dimer's size is tied to the magnitude of the ``reconstructed" nesting vector $\vec q_a \approx (0.4\pi, 0)$. Since the cobalt impurity potential is repulsive, LDOS is small at the impurity
site and large when $\vec r = (\pm a \frac12 2\pi/0.4 \pi , 0) \approx (\pm3 a, 0)$. Therefore, the size of the dimer is $6a$, close to $8a$ of Ref. \cite{DavisDimer}. In addition, the $C_4$ symmetry is broken --- even if $\Delta_{PoDW} = \Delta_{SDW}$ --- due to the different orbital components within each pocket. The charge susceptibility, therefore, peaks along the direction perpendicular to PoDW, and results in the $a$-axis orientation of the dimer.

In summary, we have shown that the anisotropic electronic dimer structure can be understood based on the ``hidden" PoDW order in parent compounds. This order induces a ``reconstructed" nesting vector $\vec q_a$, at which the charge susceptibility $\chi(\vec q_a)$ develops pronounced peaks. Furthermore, the $C_4$ symmetry is genetically broken due to the orbital components. The dimer, therefore, points along the $a$ axis of AFM order, in accordance with experiments.
More generally, our results accentuate the potential
of local probes as diagnostic tools in unraveling the patterns of various
forms of order in iron pnictides.

We thank Milan Allan, J. C. Seamus Davis, and also J. E. Hoffman for discussions and for
sharing data with us prior to publication. We thank Pierre Richard and Nachum Plonka
for comments. This work was supported in part by the IQM, under Grant No.\ DE-FG02-08ER46544 by the U.S.\ DOE, Office of Basic Energy Sciences, Division of Materials Sciences and Engineering.

%\par
%\vskip .2truein
%*******  END **********
%\vskip .2truein
%\par

%\end{widetext}
%\end{multicols}

%\bibliographystyle{apsrev}

%\bibitem {Keldish} L.\ V.\ Keldish, and Yu.\ V.\ Kopaev, Sov.\ Phys.\ - Solid Stat.\ , {\bf 6}, 2219 (1965).

\end {document}